# Measuring contextual citation impact of scientific journals

Henk F. Moed

Centre for Science and Technology Studies (CWTS)
Leiden University
PO Box 905
2300 AX Leiden
The Netherlands
Email: moed@cwts.leidenuniv.nl

**Abstract**

This paper explores a new indicator of journal citation impact, denoted as source normalized impact per paper (*SNIP*). It measures a journal's *contextual* citation impact, taking into account characteristics of its properly defined subject field, especially the frequency at which authors cite other papers in their reference lists, the rapidity of maturing of citation impact, and the extent to which a database used for the assessment covers the field's literature. It further develops Eugene Garfield's notions of a field's 'citation potential' defined as the average length of references lists in a field and determining the probability of being cited, and the need in fair performance assessments to correct for differences between subject fields. A journal's subject field is defined as the set of papers citing that journal. *SNIP* is defined as the ratio of the journal's citation count per paper and the citation potential in its subject field. It aims to allow direct comparison of sources in different subject fields. Citation potential is shown to vary not only *between* journal subject categories – groupings of journals sharing a research field – or disciplines (e.g., journals in mathematics, engineering and social sciences tend to have lower values than titles in life sciences), but also between journals *within* the same subject category. For instance, basic journals tend to show higher citation potentials than applied or clinical journals, and journals covering emerging topics higher than periodicals in classical subjects or more general journals. *SNIP* corrects for such differences. Its strengths and limitations are critically discussed, and suggestions are made for further research. All empirical results are derived from Elsevier's *Scopus*.

# 1. Introduction

The journal impact factor developed by Eugene Garfield and published by the *Institute for Scientific Information* (currently *Thomson Reuters*) in the *Journal Citation Reports (JCR)* is probably the most widely dispersed bibliometric construct. Numerous authors have discussed the potentialities and limitations of the impact factor and other measures of journal citation impact (e.g., Garfield, 1972; 1996; Glanzel and Moed, 2001).

Garfield (1979) underlined that it is improper to make comparisons between citation counts generated in different research fields, because the "citation potential" can vary significantly from one field to another. He suggested that "the most accurate measure of citation potential is the average number of references per paper published in a given field". He argued that since biochemical papers contain 30 cited references and mathematics articles 15, the citation potential in the former discipline is two times that in the latter. Moreover, variations exist in "citation characteristics as to how quickly a paper will be cited, how long the citation rate will take to peak and how long the paper will continue being cited" (Garfield, 1979, p. 248).

Garfield also emphasized that disciplinary distinctions made between fields may not always be fine enough to avoid unfair comparisons. The potential of being cited differs substantially not only between disciplines but also from one specialty to another. "Evaluation studies using citation data must be very sensitive to all divisions, both subtle and gross, between areas of research; and when they are found, the study must properly compensate for disparities in citation potential"(Garfield, 1979, p 249).

One way to overcome differences in citation potential is applying 'relative' indicators that calculate the ratio of a journal's citation impact per paper and the world citation average in the subject field the journal covers. This approach can be denoted as *target* normalization or, as Zitt and Small (2008) put it, 'cited side' normalization (Zitt and Small, 2008). Several authors proposed useful indicators based on this principle using a categorization of scientific journals into some 150 subject categories (e.g., Braun, Glänzel and Schubert (1988); Sen, 1992; Marshakova-Shaikevich, 1996; Van Leeuwen et al., 2002). The second column of Table 1 presents the base characteristics of this approach.

**Table 1: Target or 'cited side' versus source or 'citing side' normalization of journal citation impact measures**

| Steps | Target or 'cited side' normalization | Source or 'citing side' normalization |
|---|---|---|
| 1 | Calculate a journal's average citation count per paper | |
| 2 | Define the subject field covered by a journal | |
| 3 | Calculate how frequently papers in the subject field are cited by other papers (a subject field's received citation rate) | Calculate how frequently papers in the subject field cite other papers (a subject field's citation potential) |
| 4 | Correct a journal's citation count per paper for differences in received citation rates between subject fields | Correct a journal's citation count per paper for differences in citation potential between subject fields |

4An alternative approach, summarized in the right column of Table 1 is *source* or 'citing side' normalization, which corrects for differences between research fields in the frequency at which papers cite other documents, denoted as the field's 'propensity to cite' (Zitt and Small, 2008) or 'citation potential' (Garfield, 1979).

Zitt and Small's work further explores the citing-cited journal matrix and applies source normalization at the level of the citing and cited *journal*. They defined the field covered by a journal in terms of a collection of journals citing it. Their approach is related to the idea of using fractional citation counting, according to which "each citing item has a total voting strength of one, but divides that single vote equally among all references it cites"( Small and Sweeney, 1985).

During the past years, numerous other approaches to the measurement and ranking of journal impact or status were explored. Without claiming completeness, important approaches are:
- A ranking procedure similar to percentile ranking, generating rank-normalised impact factors of scientific journals based on citation analysis (e.g., Pudovkin and Garfield, 2004).
- Following Pinski and Narin (1976), application of a (variant of the) PageRank algorithm to the journal-to-journal citation network (Bollen, Rodriguez and Van de Sompel, 2005; Bergstrom, 2007; West et al., 2008; SCIMAGO).
- Following Hirsch (2005), the calculation of Hirsch Indices for scientific journals based on citations (e.g., Braun, Glanzel and Schubert, 2005).
- Development of a model for the asymptotic number of citations collected by papers published in a journal, enabling one to quantify both the typical impact and the range of impacts of papers published in a journal (Stringer, Sales-Pardo and Amaral, 2008).
- The use of data on the frequency of downloads of papers from electronic publication archives for the calculation of a journal 'usage' factor (e.g., Bollen and Van de Sompel, 2008).
- Modeling citation distributions in a journal as a negative binomial distribution, and characterizing a journal's impact by estimating the parameters of its distribution (Glanzel, 2009).

The study presented in this paper further develops the ideas by Garfield, Zitt and Small outlined above, and presents a new indicator of journal citation impact, denoted as source normalized impact per paper (*SNIP*). It measures a journal's *contextual* citation impact, taking into account characteristics of its subject field, especially the frequency at which authors cite other papers in their reference lists, the rapidity of maturing of citation impact, and the extent to which the database used for the assessment covers the field's literature. Its base principles and main characteristics are outlined in ***Section 2***. ***Appendix A1*** provides a mathematical framework and gives methodological details. ***Section 3*** illustrates the effect of using the new source normalized measure upon rankings of journals. Finally, ***Section 4*** presents a critical discussion of the new metric's strengths and limitations, compares it to other measures of journal citation impact, and makes suggestions for further research. All empirical results presented in this paper are derived from Elsevier's *Scopus*.

## 2. Base principles

*Document types included*

The methodology described in this paper aims to analyze peer reviewed research articles, and to capture citations from peer reviewed articles to other peer reviewed articles. *Scopus* uses a categorization of documents into 15 document types. In this study *articles*, *conference proceedings papers* and *reviews* are considered as fully fledged, peer-reviewed research articles. They will be denoted as 'papers' and sometimes as 'articles' throughout this article. The publication and citation counts in this study are based on these three document types only, as if all other document types were simply erased from the database.

*Citation potential*

A key concept is *citation potential*. Figure 1 illustrates how it differs from the concept of citation impact. The former indicates how frequently papers in a subject field cite other papers, and the latter how frequently a subject field's papers are cited. A subject field's citation potential is defined as the average number of cited references per paper in the subject field.

**Figure 1. Citation potential versus citation impact**

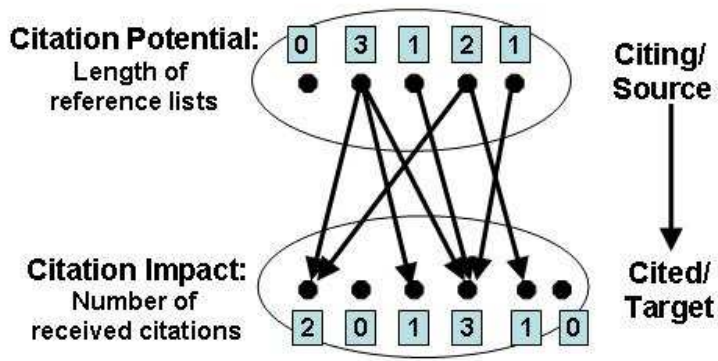

*Legend to Figure 1:* Figure 1 shows a set of 5 source articles, and their citation relationships with 6 target articles. Numbers in the upper oval indicate the number of cited references contained in each source paper. Citation potential in a set of source articles is defined as the average number of cited references per source article, which amounts in this example to 7/5 or 1.4.

*Delimitation of a journal's subject field*

A journal's subject field is defined as the collection of *papers* citing that journal. This is indicated in Figure 2. Each paper in the field cites at least one article published in the particular journal. But it is essential to realize that these papers cite other documents as well. This is shown in Figure 3. Most of these cited documents are not published in the journal itself, but in other journals or in other types of sources such as books. In fact, it is shown in Section 3 that in the reference lists of papers citing a journal the percentage of citations to the journal itself is typically only one per cent.

**Figure 2. Delimitation of a journal's subject field**

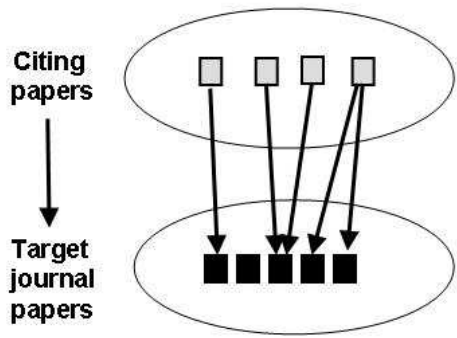

**Legend to Figure 2.** Figure 2 illustrates that the subject field of a journal (denoted as target journal) is defined as the collection of papers citing that journal. More technical details, particularly about citation time windows applied, are given in Appendix A1.

**Figure 3. Complete reference lists in papers citing the target journal**

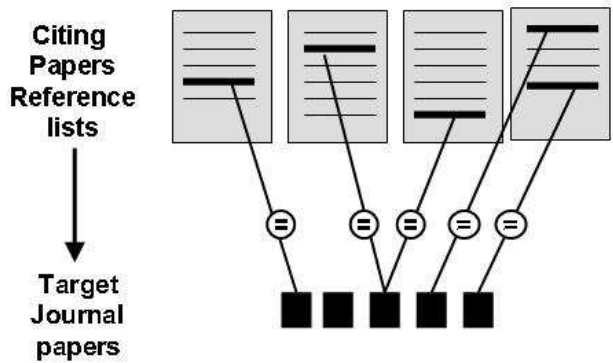

**Legend to Figure 3.** Figure 3 illustrates that the papers citing a particular (target) journal cite in their reference lists publications in other sources as well. The complete reference lists are used to calculate the citation potential in the journal's subject field.

*Database coverage and database citation potential*

The citation potential in a journal's subject field depends upon the extent to which the database covers this field. The study presented in this paper calculates indicators for journals that are processed for a particular database, *Scopus*. These are often denoted as 'source journals'. But if the papers in a field covered by a particular journal mainly cite documents published in journals or other types of sources that are *not* covered by the database, this journal will not be cited frequently, since most of the citations are directed towards documents that are not in the database. In a sense, such citations are

'lost' for target journals processed for the database. Figure 4 shows that for all papers in the 2007 *Scopus* database, about 80 per cent of cited references is published in journals or other types of sources processed for *Scopus*. This percentage is an indicator of database coverage (Moed, 2005). However, as is illustrated in Section 3 large differences exist in this coverage percentage between disciplines and subject fields.

**Figure 4. Internal coverage of the Scopus database**

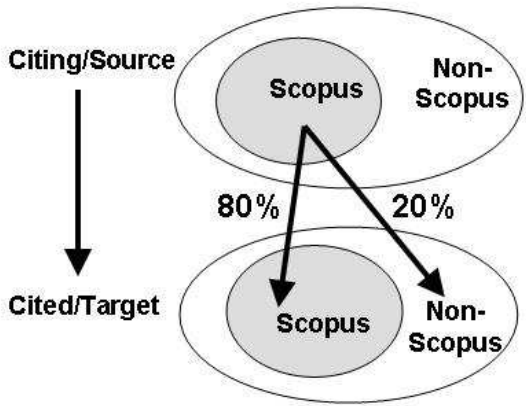

*Legend to Figure 4*. Figure 4 shows that 80 per cent of cited references in source documents in *Scopus* is published in journals or other sources processed for *Scopus*. The remaining 20 per cent is published in sources *not* processed for *Scopus*. Data are obtained from a bibliometric version of *Scopus* created at CWTS, based on raw data extracted form *Scopus* in September 2008. Citing year: 2007; cited years: 1997-2006; based on citations from articles, proceedings papers and reviews to other articles, proceedings papers or reviews.

**Figure 5. Citation potential versus database citation potential: An example**

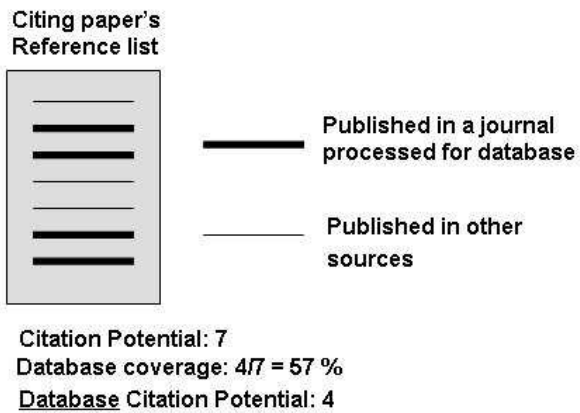

**Legend to Figure 5**. Figure 5 shows an example of a source paper with 7 cited references, four of which are published in sources processed for the database. Database coverage is defined as the percentage of such references, relative to the total number of cited references. Database Citation Potential is simply the number of cited references published in sources processed for the database.

Therefore, citation potential must take into account the extent to which the database covers a subject field. A new parameter is defined, denoted as *database* citation potential. It does *not* count the *total* number of cited references in a field's papers, but the number of cited references *published in journals processed for the database*. Figure 5 shows an example of how database citation potential is calculated.

*Citation and publication windows*

The citation impact indicator presented in this paper is based on citations given in a fixed *citing* year (2007) to a journal's papers published in the *three* preceding years (2004-2006). Compared to the *Thomson Reuters'* journal impact factor published in the *Journal Citation Reports* (***JCR***), the citation time window of this new metric is one year longer, giving on average a journal's impact more time to mature. This is particularly useful in disciplines in which citation impact matures slowly (Moed, Van Leeuwen and Reedijk, 1998) such as *mathematics* (Rousseau, 1988) and parts of *engineering*, *social sciences* and *humanities.* An important cause of slowly maturing impact in a field is the existence of a long publication delay – i.e., the time period between a paper's date of submission to a journal and its formal publication date.

*Examples: Database citation potential in three subject fields*

Figure 6 shows the distribution of the number of cited references – published in journals processed for the database – between papers citing one of three journals: *Inventiones Mathematicae*, *Molecular Cell*, and *Journal of Electronic Materials*. The database citation potential in the subject fields covered by these three journals is 2.86, 22.21 and 6.87, respectively.

*Raw impact per paper (RIP)*

The source normalized impact per paper (*SNIP*) of a journal is a ratio. This paragraph describes the *numerator* in that ratio: the average number of citations per paper published in a journal, denoted as raw impact per paper (*RIP*). As outlined above, citations are counted that are given in a fixed (citing) year (in this study 2007) to papers published in the journal during the three preceding years (2004-2006). Table 2 clarifies the distinction between a journal's raw impact per paper and the database citation potential in its subject field.

**Table 2.    Raw Impact per Paper versus Database Citation Potential**

| *A journal's raw impact per paper (RIP)* | *Database Citation Potential of a journal's subject field* |
|---|---|
| Average number of citations (from whatever source journals in the database) received by 1-3 year old papers published in the target journal | Average number of 1-3 year old cited references (published in whatever source journal processed for the database) contained in papers citing the target journal |

**Figure 6. Distribution of the number of cited references in papers citing three journals**

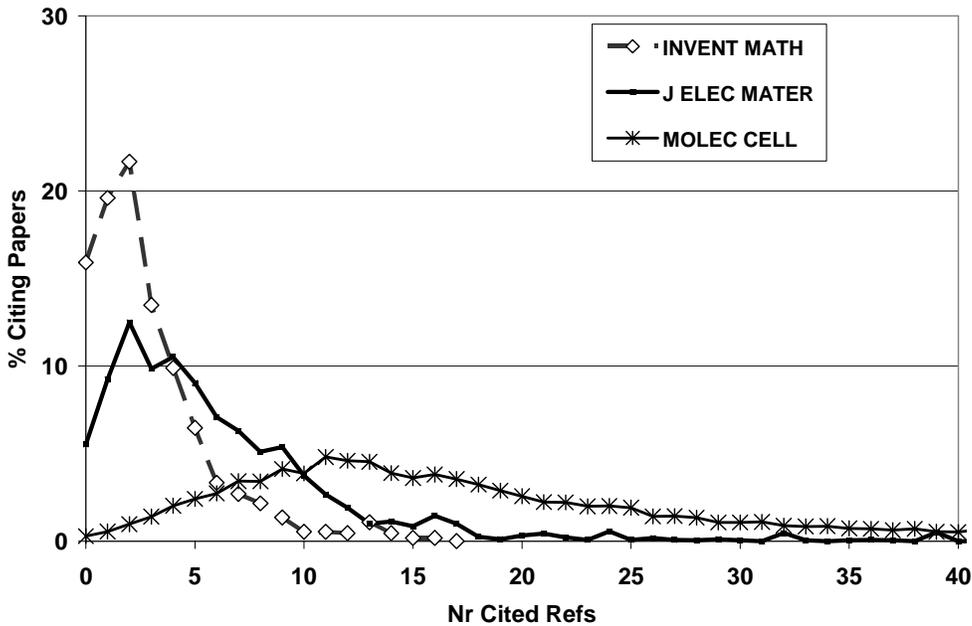

**Legend to Figure 6**: Citing year: 2008; Cited years: 2005-2007. % Citing papers with > 40 cited references are not displayed. Basic data on the three journals:

| Abbreviation | Full title | Database citation potential in subject field | Relative database citation potential in subject field |
|---|---|---|---|
| INVENT MATH | Inventiones Mathematicae | 2.86 | 0.42 |
| J ELEC MATER | Journal of Electronic Materials | 6.87 | 1.00 |
| MOLEC CELL | Molecular Cell | 22.21 | 3.23 |

If one selects all papers in the database citing in 2007 at least one article published in *Inventiones Mathematicae*, and if one counts in the cited reference list in each citing paper the number of cited references published during the three preceding years in journals processed for the database, the mean of this number over all citing papers amounts to 2.86. Note that the database citation potential in the field of *Molecular Cell* is almost one order of magnitude higher.

*Relative citation database potential*

*Journal of Electronic Materials* is the *median* journal in the database in terms of the database citation potential (*DCP*) of its subject field. In other words: 50 per cent of journals in the database has a *DCP* value above, and another 50 per cent below 6.87, the *DCP* of this journal (see Figure 6). Therefore, it is used as a *DCP* normalization factor. The *DCP* of any journal is divided by that of this median journal. This ratio is called Relative Database Citation Potential (*RDCP*). By definition, *RDCP* of the median journal equals one. 50 per cent of journals has a *RDCP* above one.

*Source normalized impact per paper (SNIP)*

The definition of a journal's source normalized impact per paper is presented in Table 3 below.

**Table 3.    Definition of a journal's source normalized impact per paper**

| A journal's Source Normalized Impact per Paper (*SNIP*) <br> = <br> Raw Impact per Paper published in the journal (*RIP*) <br> ÷ <br> Relative Database Citation Potential (*RDCP*) in the journal's subfield |
|---|

For journals in subject fields in which the average length of reference lists – corrected for citation time window and database coverage – is equal to that for the *median* journal in terms of its subject field's citation potential, the relative database citation potential equals one, and the new indicator, *SNIP*, equals *RIP*, the raw impact per paper. But in subject fields with a higher citation potential, which for instance is the case for many journals in *biochemistry and molecular biology*, *SNIP* is lower than *RIP*, whereas in fields such as *mathematics*, in which citation potentials are generally lower, *SNIP* tends to be higher than *RIP*. By using the database citation potential in the median subject field as normalization factor, half of the journals go up in *SNIP* compared to *RIP*, and the other half goes down, depending upon whether the value of the relative citation potential in their subject fields is below or above one. This is illustrated in Figure 7.

**Figure 7.   Ratio of source normalized (*SNIP*) and raw (*RIP*) impact per paper as a function of relative database citation potential**

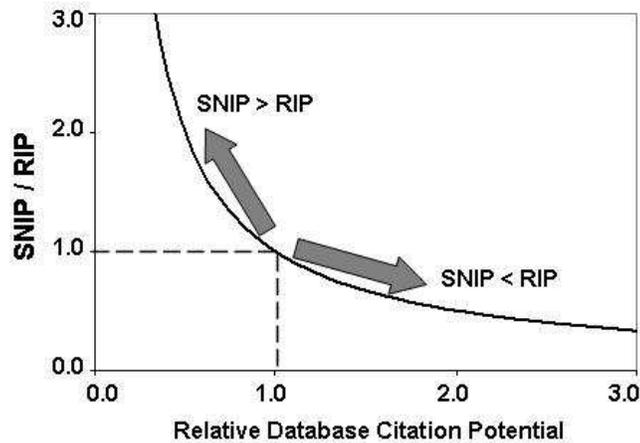

*Legend to Figure 7*: *SNIP*: Source normalized impact per paper; *RIP*: Raw impact per paper. The figure shows that if the relative database citation potential in a journal's subfield has a value of one – i.e., the database citation potential equals that for the median journal in the database – , the journal's source normalized impact per paper equals its raw impact per paper. If in a journal's subject field the relative database citation potential is below one, *SNIP* is higher than *RIP*, while if this potential is above one, *SNIP* is lower than *RIP*.

*Effect of source normalization upon citation impact distributions for all SCOPUS journals*

**Figure 8** and **Table 4** give an overall impression of the effect of source normalization upon the distribution of citation impact between all around 17,000 journals in the *Scopus* database.

**Table 4.** Statistics of the distribution of raw (*RIP*) and source normalized (*SNIP*) impact per paper across all source journals in the database

| *Indicator* | *N* | *Mean* | *Std* | *Skewness* | *P25* | *P50* | *P75* | *P90* | *P99* |
|---|---|---|---|---|---|---|---|---|---|
| SNIP | 17,000 | 0.81 | 1.12 | 16.03 | 0.17 | 0.52 | 1.10 | 1.77 | 4.68 |
| RIP | 17,000 | 1.04 | 2.10 | 25.26 | 0.13 | 0.48 | 1.32 | 2.51 | 7.59 |

*Legend to Table 4*: *SNIP*: Source normalized impact per paper. *RIP*: Raw impact per paper. The distributions are characterized by their percentile values. For instance, P50, the 50th percentile (i.e. the median) of the *SNIP* distribution amounts to 0.52. This means that 50 per cent of the 17,000 journals have a *SNIP* value up or below 0.52. P99 for *SNIP* and *RIP* are 4.68 and 7.59, respectively. This means that one percent of journals (about 1,700 journals) have a *SNIP* value above 4.68, and one per cent a *RIP* value above 7.59.

Table 4 and Figure 8 clearly show that the distribution of the source normalized impact per paper (*SNIP*) across journals is more concentrated than that based on raw impact per paper (*RIP*). In fact, the standard deviation of the *RIP* distribution is almost two times that for the *SNIP* distribution: 2.10 versus 1.12.

**Figure 8.** Distribution of raw (*RIP*) and source normalized (*SNIP*) impact per paper across all source journals in the database

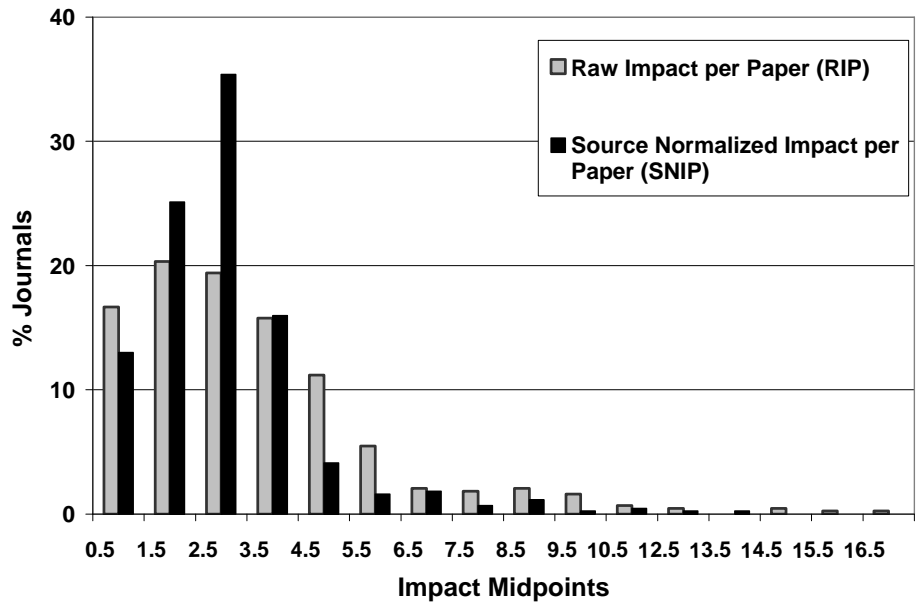

**Legend to Figure 8**: The horizontal axis gives midpoints either for a journal's source normalized or raw impact per paper. For instance, midpoint 2.5 comprises all journals with impact values between 2.0 and 3.0. Figure 8 shows that slightly less than 20 per cent of *Scopus* journals has a raw impact per paper between 2.0 and 3.0. The *SNIP* distribution shows more concentration than that for *RIP*. It is less skewed to the right. The highest scores tend to be lower than those for *RIP*. About 35 per cent of journals has a *SNIP* value between 2.0 and 3.0.

## 3. Results

**Table 5** gives an overview of the indicators calculated in this paper. **Table 6** presents *SNIP* values and the other indicators listed in Table 5 for a number of journals from various journal subject categories, using a classification implemented in *Scopus* of journals into about 300 categories. Although this table reveals differences in raw impact per paper and citation potential *between* subject categories, its primary aim is to show such differences among journals *within* subject categories.

It shows typical examples of *pairs* of journals with statistically similar *SNIP* values, revealing substantial differences in relative database citation potential (*RDCP*) in the field covered by a journal and in a journal's raw impact per paper (*RIP*). In this way, Table 6 shows how the source normalization procedure described in the previous section brings together journals with very different positions in a ranking based on their raw impact per paper. Moreover, from the titles of these journals one obtains a rough impression of the differences in covered topics between the journals. Interesting results are the following.

- The raw impact per paper (*RIP*) in *Journals of Gerontology A – Medical and Biological Sciences* is 35 per cent higher than that of *Journals of Gerontology B – Psychological and Social Sciences* (3.66 against 2.72), but the former's *source normalized* impact per paper (*SNIP*) is 22 per cent lower than that of the latter (1.81 versus 2.31). In fact, the relative database citation potentials (*RDCP*) in the subject fields covered by these two journals are 2.02 and 1.17, respectively. It is plausible to assume that this difference reflects differences in citation characteristics between medical-biological sciences on the one hand, and social sciences on the other.

- The journal pairs in the subject categories *Algebra and Number Theory* and *Applied Mathematics* do not only show low relative database citation potentials compared to those in other journal categories, but also large differences *within* each pair. Focusing on the former category, *Journal of Logic and Algebraic Programming* has a *RDCP* of 0.95, which is more than twice that for *Journal of Differential Geometry* (0.45). Their *SNIP* values are almost identical (1.97 versus 1.98). Secondary analysis reveals that the latter journal is mainly cited from purely mathematical journals such as *Transactions of the American Mathematical Society*, *Geometriae Dedicata*, and *Advances in Mathematics*, whereas the former is predominantly cited from computer science sources, especially *Lecture Notes in Computer Science*, *Theoretical Computer Science* and *Lecture Notes in Artificial Intelligence.*

- W*ithin* a molecular-biological approach differences exist among research *objects*. *Insect Biochemistry and Molecular Biology* and *Plant Molecular Biology* reveal different relative citation potentials (1.70 against 2.55). Although the raw impact per paper of the latter is 50 per cent higher than that of the former (4.27 versus 2.84), the *SNIP* values of the two journals are identical (1.67).

**Table 5 Journal indicators calculated in this paper**

| *Indicator* | *Technical details* | *Significance* |
|---|---|---|
| Nr Papers | Number of articles, reviews and proceedings papers published in a journal in the 3 years preceding the year of analysis | Indicates the number of peer-reviewed documents published in a journal. |
| % Reviews | % Papers published in a journal and labeled as reviews in the database | Review papers tend to be cited more frequently than other types; journals publishing reviews tend to have higher citation impact. |
| Raw impact per paper (*RIP*) | Number of citations in year of analysis to a journal's papers published in 3 preceding years, divided by the number of a journal's papers in these three years | Corrects for differences in sizes of annual volumes. Is similar to Thomson's JCR impact factor but is based on citations to papers published during 3 preceding years (instead of 2); 'free' citations to 'non-citable' items are not included; only citations in articles that are 'peer reviewed' are counted. |
| *C*itation potential in the journal's subject field | Mean number of 1-3 year old cited refs per paper citing a journal (e.g., cited references contained in a 2007 paper, and themselves published during 2004-2006). | Indicates how frequently papers in a journal's subject field cite other papers published in 3 preceding years. The higher this number, the higher is for 1-3 year old papers in the journal's subject field the probability of being cited |
| Database coverage of a journal's subject field | For papers in a journal's subject field: % 1-3 year old cited references *published in journals processed for the database* | Indicates how frequently papers in a journal's subject field cite other papers *published in journals that were processed for the database*. It is a measure of the extent to which the database covers the field. |
| *Database* citation potential in a journal's subject field | Mean number of 1-3 year old references per paper citing the journal *and published in journals processed for the database* | Indicates how frequently papers in a journal's subject field cite 1-3 year old other papers published in journals that were processed for the database |
| *Relative* database citation potential in a journal's subject field (*RDCP*) | Database citation potential of a journal's subject field divided by that for the median journal in the database | According to this normalization, the median journal in terms of database citation potential in its subject field has a value of one. Biochemical journals tend to have a value above one, and mathematical titles below one. |
| Source normalized impact per paper: (*SNIP*) | Ratio of a journal' raw impact per paper (*RIP*) and the relative database citation potential (*RDCP*) in the subject field covered by the journal | For journals covering subject fields in which the relative database citation potential (*RDCP*) equals one, *SNIP* equals *RIP*. For biochemical journals *SNIP* values tend to be lower than their *RIP* scores, and for mathematical periodicals higher. |
| % Journal self citations | % Citations to a journal, given in papers published in the journal itself | Indicates the fraction of a journal's raw impact per paper that is generated by the journal itself. |
| % Cited refs in subfield to journal | % Cited references in a journal's subject field published in the journal itself | Indicates the relative frequency at which papers in a journal's subject field cite that journal in their reference lists. |

**Table 6.** *SNIP* and related indicators for selected pairs of journals

| Journal | Nr Papers (2004-2006) | % Reviews | SNIP | Raw Impact per paper (RIP) | Database Coverage (%) | Database Citation Potential | Relative Database Citation Potential | % Journal self citations | % Citations in subfield to journal |
|---|---|---|---|---|---|---|---|---|---|
| *Accounting* | | | | | | | | | |
| Mathematical Finance | 91 | 2.2 | *2.93* | 1.26 | 62 | 2.96 | 0.43 | 13.9 | 0.96 |
| Financial Management | 65 | 29.2 | *2.55* | 1.78 | 69 | 4.80 | 0.70 | 11.2 | 0.70 |
| *Acoustics and Ultrasonics* | | | | | | | | | |
| J Vibration & Acoustics | 229 | 0.0 | *1.82* | 0.92 | 79 | 3.46 | 0.50 | 16.2 | 1.23 |
| Ultrasonics & Sonochemistry | 241 | 3.7 | *2.03* | 2.58 | 89 | 8.71 | 1.27 | 17.4 | 0.87 |
| *Aging* | | | | | | | | | |
| J Gerontol - A Biol & Med Sci | 559 | 13.6 | *1.81* | 3.66 | 90 | 13.89 | 2.02 | 6.9 | 0.17 |
| J Gerontol - B Psych & Soc Sci | 274 | 7.7 | *2.31* | 2.72 | 79 | 8.07 | 1.17 | 11.7 | 0.36 |
| *Algebra and Number Theory* | | | | | | | | | |
| J Logic and Algebr Program | 75 | 5.3 | *1.97* | 1.87 | 68 | 6.52 | 0.95 | 4.3 | 0.46 |
| J Differential Geometry | 114 | 3.5 | *1.98* | 0.89 | 70 | 3.07 | 0.45 | 8.9 | 0.61 |
| *Analytical Chemistry* | | | | | | | | | |
| J Chromatogr A | 3,872 | 3.2 | *1.56* | 3.62 | 93 | 15.89 | 2.31 | 22.5 | 0.57 |
| J Electroanalyt Chem | 1,227 | 0.8 | *1.62* | 2.67 | 92 | 11.32 | 1.65 | 10.5 | 0.29 |
| *Anatomy* | | | | | | | | | |
| Clin Anatomy | 331 | 8.5 | *0.96* | 0.85 | 87 | 6.06 | 0.88 | 12.8 | 0.70 |
| Cells Tissues Organs | 177 | 5.1 | *0.99* | 2.39 | 95 | 16.53 | 2.41 | 4.7 | 0.12 |
| *Applied Mathematics* | | | | | | | | | |
| Int J Nonlinear Sci & Numer Simulatation | 190 | 1.6 | *2.13* | 4.24 | 95 | 13.68 | 1.99 | 15.4 | 0.93 |
| Commun Partial Differential Equations | 215 | 1.4 | *2.13* | 1.06 | 75 | 3.41 | 0.50 | 5.3 | 0.43 |
| *Atomic and Molecular Physics, and Optics* | | | | | | | | | |
| Chemphyschem | 883 | 5.4 | *1.42* | 3.39 | 86 | 16.43 | 2.39 | 4.2 | 0.16 |
| Optics & Laser Technology | 307 | 1.0 | *1.42* | 0.90 | 85 | 4.34 | 0.63 | 12.0 | 1.01 |
| *Atomic and Molecular Physics, and Optics* | | | | | | | | | |
| J Mol Spectrosc | 526 | 0.4 | 1.15 | 1.14 | 79 | 6.79 | 0.99 | 35.7 | 1.74 |
| J Nanoparticle Res | 209 | 1.9 | 1.20 | 2.26 | 83 | 12.98 | 1.89 | 6.5 | 0.21 |
| *Aquatic Science* | | | | | | | | | |
| Aquatic Toxicol | 519 | 2.3 | *1.88* | 3.45 | 92 | 12.60 | 1.84 | 15.3 | 0.55 |
| Continental Shelf Res | 428 | 5.8 | *1.89* | 2.06 | 84 | 7.51 | 1.09 | 14.2 | 0.65 |
| *Behavioral Neuroscience* | | | | | | | | | |
| Behaviour | 248 | 2.8 | *1.21* | 1.78 | 86 | 10.07 | 1.47 | 9.3 | 0.27 |
| Physiology & Behavior | 886 | 7.5 | *1.24* | 2.93 | 93 | 16.18 | 2.36 | 8.1 | 0.18 |

| Table 6. SNIP and related indicators for selected pairs of journals (continued) | | | | | | | | |
|---|---|---|---|---|---|---|---|---|
| *Journal* | *Nr Papers* | *% Reviews* | *SNIP* | *Raw impact per paper (RIP)* | *Database Coverage (%)* | *Database Citation Potential* | *Relative Database Citation Potential* | *% Journal self citations* | *% Cited refs in subfield to journal* |
| *Biochemistry* | | | | | | | | | |
| Insect Biochem & Molec Biol | 335 | 2.4 | *1.67* | 2.84 | 92 | 11.68 | 1.70 | 13.3 | 0.37 |
| Plant Molec Biol | 577 | 5.0 | *1.67* | 4.27 | 95 | 17.54 | 2.55 | 4.8 | 0.08 |
| *Bioengineering* | | | | | | | | | |
| Bioresource Technol | 829 | 2.5 | *2.52* | 3.33 | 89 | 9.06 | 1.32 | 10.5 | 0.44 |
| Biomaterials | 2109 | 2.9 | *2.99* | 6.53 | 94 | 14.98 | 2.18 | 9.3 | 0.30 |
| *Cardiology and Cardiovascular Medicine* | | | | | | | | | |
| Arteriosclerosis, Thrombosis & Vascular Biol | 1,021 | 3.4 | *2.46* | 6.45 | 94 | 18.03 | 2.63 | 4.6 | 0.08 |
| J Vascular Surg | 1,190 | 13.4 | *2.50* | 4.15 | 93 | 11.40 | 1.66 | 17.0 | 0.58 |
| *Ecology* | | | | | | | | | |
| Ecology | 998 | 7.8 | *3.46* | 5.22 | 87 | 10.36 | 1.51 | 8.0 | 0.23 |
| Ecology Letters | 414 | 22.7 | *4.52* | 8.63 | 87 | 13.11 | 1.91 | 5.2 | 0.22 |
| *Engineering (miscellaneous)* | | | | | | | | | |
| Combustion Sci & Technol | 284 | 7.0 | *1.60* | 1.28 | 80 | 5.49 | 0.80 | 7.4 | 0.40 |
| Nanotechnology | 2,093 | 3.2 | *1.66* | 3.27 | 84 | 13.56 | 1.98 | 12.7 | 0.69 |
| *Plant sciences* | | | | | | | | | |
| Field Crops Res | 411 | 1.0 | 1.99 | 2.03 | 78 | 7.00 | 1.02 | 14.8 | 0.70 |
| Plant Cell | 780 | 6.9 | 3.51 | 10.27 | 96 | 20.08 | 2.92 | 8.5 | 0.16 |
| *General / Multidisciplinary* | | | | | | | | | |
| Nature | 3,966 | 11.9 | 7.62 | 19.02 | 90 | 17.13 | 2.49 | 1.3 | 0.02 |
| Science | 4,477 | 24.5 | 6.26 | 15.40 | 89 | 16.90 | 2.46 | 1.1 | 0.02 |

- Journals covering *emerging* topics tend to have higher citation potentials than journals publishing more papers in 'classical' topics, or in more general journals covering a wide range of topics. Good examples illustrating this are *Journal of Nanoparticle Research* versus *Journal of Molecular Spectroscopy* in the subject category *Atomic and Molecular Physics, and Optics*, the journal *Ultrasonics and Sonochemistry* compared to *Journal of Vibration and Acoustics* in the subject category *Acoustics and Ultrasonics*, and the pair *Nanotechnology* versus *Combustion Science & Technology* in the category *Engineering*.

- The subject category *Behavioural Neuroscience* is rather heterogeneous in terms of topics and approaches. Table 6 lists two journals from this category. *Behaviour* seems to publish mainly research on animals. The journals most frequently citing this periodical are in fact: *Animal Behavior*, *Ethology*, and *Behavioural Ecology and Sociobiology*. *Physiology & Behavior* is more focusing on human brain research, and is frequently cited from journals such as *Behavioural Brain Research*, *Hormones and Behavior*, and *American Journal of Physiology*. The subject fields covering the two listed journals have different citation potentials (1.47 against 2.36) and raw impacts per published paper (1.78 versus 2.93). Correcting for these differences, their *SNIP* values are almost equal (1.21 against 1.24).

- The journal pairs from the subject categories *Anatomy* (*Clinical Anatomy* versus *Cells Tissues Organs*) and *Cardiology and Cardiovascular Medicine* (*Arteriosclerosis, Thrombosis, and Vascular Biology* versus *Journal of Vascular Surgery*) illustrate that *clinical* journals tend to have lower database citation potentials than more *basic* oriented medical-biological periodicals. In physical sciences, basic journals tend to show higher citation potentials than applied journals. A typical example is the pair *Chemphyschem* and *Optics and Laser Technology*.

- Journals publishing letters or short communications tend to show higher citation potentials than 'normal' journals. Apparently, authors citing letter journals are more focused on the recent (i.c., 1-3 year old) literature. The relative database citation potential for *Ecology Letters* is about 25 per cent higher than that of *Ecology* (1.91 versus 1.51). It must be noted that the percentage of review articles is higher for the former than it is for the latter (22.7 versus 7.8 per cent).

- The last two rows in Table 6 present indicators for two general or multidisciplinary journals, *Nature* and *Science*. The methodology described in this study enables one to calculate a source normalized impact also for this type of journals. A *SNIP* value is found of 7.62 for *Nature* and 6.26 for *Science*.

## 4. Discussion and conclusions

The aim of this Section is to critically evaluate the potentialities of the proposed indicator, but at the same time to be aware of its limitations. There is no single 'perfect' indicator of journal performance. The scholarly communication system is highly complex, citations constitute one of its representations – though a most valid and useful one – and journal performance is a multi-dimensional concept that cannot be expressed in any single measure. The adequacy of a journal impact measure is related to the type of use made of it, and the type of research question addressed. A particular indicator may be appropriate in one context, and less appropriate in another.

Generally speaking, this metric assesses a journal's citation impact '*in context*'. This context is determined by reference practices in peer reviewed articles in the journal's subject field and by the extent to which the database covers this field. Strong points of the *SNIP* metric are the following.

- Delimitation of a journal's subject field does *not* depend upon some pre-defined categorization of journals into subject categories, but is entirely based on citation relationships. It is carried out on a (citing) paper-by-paper basis, rather than on a (citing) journal-by-journal basis.
- The delimitation is 'tailor-made'. A subject field can be defined accurately even when general or multi-disciplinary journals covering several fields rather than one play an important role in it.
- The new metric corrects for differences in referencing practices between subject fields, especially the frequency at which authors cite other papers, and the rapidity of maturing of citation impact.
- In addition, it corrects for differences in database coverage between subject fields: lower database coverage leads to lower database citation potentials, which tends to lead to higher *SNIP* values compared to the raw impact per paper indicator.
- It does not only correct for differences *between* journal subject categories (i.e., groupings of journals into a few hundred research subfields), but also between journals covering distinct topics, approaches or research objects *within* a journal subject category.
- *SNIP* is based on citations from peer-reviewed papers to other peer-reviewed papers. This makes it less sensitive to manipulation and strategic behavior, especially by journal editors. 'Free' citations to non-citable documents (Moed and van Leeuwen, 1996) and 'editorial' self citations (Reedijk and Moed, 2008) are *not* included.
- It enables the calculation of sensible citation impact measures of general or multidisciplinary journals such as the journals *Nature* or *Science*.
- It is moderately sensitive to variations in the length time windows used in indicator calculation and field delimitation, at least compared to the variability between years in which citations are counted (see Appendix A1).

However, important points that should be kept in mind on the interpretation of source normalized impact per paper (*SNIP*) are:

- Contrary to a target normalized indicator (e.g., van Leeuwen et al., 2008), *SNIP* does *not* correct for differences in the fraction of *review* articles published in a journal. Similar to the raw impact per paper indicator, *SNIP* values tend to be higher for journals publishing reviews. The percentage of reviews in a journal is an important additional indicator that should be used when interpreting *SNIP*.

- The categorization of a journal's documents into document types affects the values of *SNIP* in the same measure as it influences *RIP*. A breakdown of a journal's papers into document types in the database may provide useful background information on *SNIP*.
- Although the impact or quality of journals used for publication is an aspect of research performance in its own right, journal impact factors should not be used as surrogates of citation impact of individual papers or research group publication oeuvres (Garfield, 1996; Seglen, 1994; 1997) This is true both for the raw and the source normalized impact per paper.
- The higher the percentage of journal self citations, the more the journal's indicators are determined by citations from – and cited reference characteristics within – the journal itself. When interpreting *SNIP* or *RIP* values, this percentage provides relevant background knowledge.
- While the new indicator does correct for differences in citation potential between subject fields – as expressed in the length of cited reference lists in papers covering the field – , it does *not* take into account the growth of the literature in a field, nor the extent to which papers in a field are cited from other fields (Zitt and Small, 2008).

The following issues await further research.

- The field delimitation explored in this paper is a *first-order* delimitation. More sophisticated methodologies based on citation analysis are feasible, for instance, those involving an iterative process in which a next step could be adding the papers citing with a particular strength the documents that are cited by the articles published in the journal.
- Papers belonging to a subject field but *not citing* a journal (in the time window applied) are by definition not included in a journal's field, and this could cause a *bias*. Although this bias is reduced by applying in the field delimitation a time period of cited years that is much longer than that used in the actual indicator calculation, and although the analyses presented in the Appendix suggest that the effect of variations in citation time windows upon *SNIP* values is relatively small, the effect of the missing 'non-citing' papers upon the values obtained for *SNIP* could be examined in more detail.
- More qualitative research could examine the extent to which rankings of journals based on the new indicator correlate with the opinion of peers on the quality of journals in their fields.

**APPENDIX A1: Mathematical model and methodological details**

*A note on terminology*

The database used in this paper, Elsevier's *Scopus*, does not only include scientific-scholarly journals, but also conference proceedings, books and trade journals. However, in this paper all sources processed for the database are labeled as journals.

*Subject field delimitation*

The research field covered by a journal is defined as the collection of articles citing that journal within a particular time window specified below. This collection will be labeled as a journal's *subject field* throughout this paper.

More specifically, as a rule, a journal's field in a particular year is defined as the set of papers published in that year, and citing at least one paper published in that journal during th*e ten* preceding years. It needs emphasizing that the time window applied in the calculation of citation impact indictors is different from that used to delimit a subject field: the latter takes into account only citations in a fixed citing year to papers published during the first *three* preceding years. In the mathematical framework presented below the citation and publication time windows are not specified.

*Citation potential and database coverage*

If the set of articles citing a particular journal $j$ contains $m$ articles, $^1a_j \ldots {}^m a_j$, and if $^i r_j$ indicates the number of cited references contained in article $^i a_j$, the Citation Potential $R_j$ in the journal's subject field is defined as

$$R_j = \frac{\sum_{i=1}^{m} {}^i r_j}{m} \quad (1)$$

$R_j$ is the arithmetic mean of the number of cited references contained in papers citing a particular journal.

This study calculates for a paper $i$ in journal $J$'s subject field the number of cited references contained in paper $i$ and published in sources that are processed for the database, denoted as $^i r_j^{db}$. Similar to formula (1), the citation potential of source journals processed for the database, $R_j^{db}$ can be defined as

$$R_j^{db} = \frac{\sum_{i=1}^{m} {}^i r_j^{db}}{m} \quad (2)$$

This quantity is labeled as the database citation potential throughout this paper. If $f_j$ denotes for journal $J$'s subject field the fraction of cited references that is published in sources processed for the database, it follows that

$$R_j^{db} = f_j \cdot R_j \quad (3)$$

$f_j$ can be interpreted as an indicator of the internal database coverage of $J$'s subject field (Moed, 2005).

*Raw and source normalized impact per paper*

If $C_j$ denotes the number of cites in a particular year to papers published in $J$ during the three preceding years, and $A_j$ the number of papers published in these years, the *raw* citation impact per paper (*RIP*) of $J$ is defined as the ratio of these two quantities:

$$RIP_j = \frac{C_j}{A_j} \quad (4).$$

Let N be the number of source journals in the database, and $M^{db}$ be the median database citation potential between the subject fields of all source journals in the database, i.e.,

$$M^{db} = \text{median in} \{R_j^{db}\}, j=1 \text{ to } N \quad (5)$$

The measure proposed in this study, the source normalized impact per paper (*SNIP*) of journal *J* is defined as follows:

$$SNIP_j = \frac{RIP_j}{\dfrac{R_j^{db}}{M^{db}}} \quad (6)$$

The ratio $\dfrac{R_j^{db}}{M^{db}}$ can be denoted as a *relative* database citation potential. $SNIP_j$ is therefore a normalized citation-per-paper ratio. It expresses this ratio per 'unit' of database citation potential in the subject field covered by the journal.

*Effects of variations in publication and citation time windows*

**Table A1** analyses the sensitivity of the *SNIP* indicator for changes in the publication and citation time windows applied. The following three variations are examined:

- As outlined above, a journal's subject field is defined as the set of papers citing in a particular year (2007 in this study) at least one paper published in the journal during the *ten* preceding years – in this case during the time period 1997-2006. How does *SNIP* change if this time period is shortened to the *three* preceding years – 2004-2006, the very same time period as that applied in the calculation of the indicators?
- The impact indicators are based on citations in a fixed year (2007) to a journal's papers published in the *three* preceding years (2004-2006). How does *SNIP* vary if one counts only citations to papers from the *two* preceding years (2005-2006)?
- The indicators presented in this study are calculated for one single, fixed citation year (2007). How does *SNIP* change if one calculates indicators for an earlier fixed citing year, but adjusting the time windows of cited years in field delimitation (1996-2005) and indicator calculation (2003-2005) accordingly?

The results are presented in Table A1. It analyses differences between the default configuration (fixed citing year 2007; cited years in field delimitation 1997-2006; and cited years in indicator calculation 2004-2006) and the variant. If $V_d$ denotes the value of an indicator in the default configuration, and $V_v$ that of the variant, the difference *DIFF* is defined as follows:

$$DIFF = \left| 100 \cdot \frac{V_v - V_d}{\frac{V_v + V_d}{2}} \right|.$$

Table A1 presents the mean and median value of *DIFF* for two indicators – the raw (*RIP*) and source normalized impact per paper (*SNIP*) – over all journals covered by the database, and for a subset of 'bigger' journals, i.e., journals publishing at least one hundred papers per year during the time period 2004-2006. Standard deviations tend to be in the same order of magnitude as the means.

A general conclusion is that the variations in *SNIP* are only slightly higher than those in *RIP*, and for the set of all journals slightly higher than for that of 'bigger' journals publishing at least 100 papers. Moreover, among the three variations, changing the fixed citing year has the largest effect both on *RIP* and on *SNIP*, and shortening the time window for cited years in journals' subject field delimitation the smallest.

To be specific, as regards the latter variation – that has no effect on *RIP* – Table A1 shows that for half of the journals *SNIP* varies with at most 7.6 per cent. Considering only bigger journals, this percentage is slightly lower. Changing the fixed citing year causes for half of the journals a difference of at most 10.1 per cent in *RIP*, and of at most 11.7 per cent in *SNIP*. On the one hand, this finding suggests that the effect of changes in the cited year time period in field delimitation is relatively small. On the other hand, it provides a ground to calculate moving two or three year averages – i.e., calculating an average for scores for subsequent citing years 2007, 2006 and possibly 2005– , rather than scores for one single citing year.

**Table A1. The effect of variations in the citation and publication time window upon *SNIP* and impact per paper**

| *Type of variation* | *Differences between 'default' and 'variant'* | | | |
|---|---|---|---|---|
| | *All journals* | | *Journals with >=100 papers* | |
| | *Mean* | *Median* | *Mean* | *Median* |
| *Raw impact per paper (RIP)* | | | | |
| Cited years in field delimitation 2004-2006 in stead of 1997-2006 | 0.0 % | 0.0 % | 0.0 % | 0.0 % |
| Cited years in indicator calculation 2005-2006 in stead of 2004-2006 | 9.0 % | 4.5 % | 7.7 % | 4.7 % |
| Citing year 2006 in stead of 2007 | 17.4 % | 10.1 % | 12.7 % | 7.3 % |
| *Source normalized impact per paper (SNIP)* | | | | |
| Cited years in field delimitation 2004-2006 in stead of 1997-2006 | 7.8 % | 7.6 % | 6.4 % | 5.1 % |
| Cited years in indicator calculation 2005-2006 in stead of 2004-2006 | 12.3 % | 9.0 % | 10.2 % | 8.2 % |
| Citing year 2006 in stead of 2007 | 19.7 % | 11.7 % | 14.7 % | 8.6 % |

**References**

Bergstrom, C. (2007). Eigenfactor: Measuring the value and prestige of scholarly journals. College & Research Libraries News, 68(5). Retrieved April 24, 2008, from ww.ala.org/ala/acrl/acrlpubs/crlnews/backissues2007/may07/eigenfactor.cfm

Bollen J, Rodriguez MA , and Van De Sompel, H. (2006). Journal Status. *Scientometrics,* 69, 669-687.

Bollen J, and Van de Sompel H.(2008).Usage impact factor: the effects of sample characteristics on usage-based impact metrics. Journal of the American Society for Information Science and technology, 59, 1 14.

Braun, T., Glänzel, W., and Schubert, A. (1988). World flash on basic research – The newest version of the facts and figures on publication output and relative citation impact of 100 countries 1981–1985. *Scientometrics*, 13, 181–188.

Braun T, Glanzel W, and Schubert A. (2005). A Hirsch-type index for journals. The Scientist, 19, 8.

Garfield, E. (1972). Citation Analysis as a tool in journal evaluation. *Science,* 178, 471–479.

Garfield, E. (1979). *Citation Indexing. Its theory and application in science, technology and humanities*. New York: Wiley.

Garfield, E. (1996). How can impact factors be improved? *British Medical Journal*, 313, 411–413.

Glänzel, W., and Moed, H.F. (2002). Journal impact measures in bibliometric research. *Scientometrics*, 53, 2, 171–194.

Glanzel, W. (2009). The multi-dimensionality of journal impact. *Scientometrics*, 78, 355-374.

Hirsch, JE. (2005).An index to quantify an individual's scientific research output. Proceedings of the National Academy of Sciences. 102, 16569–16572.

Marshakova-Shaikevich, I. (1996). The standard impact factor as an evaluation tool of science and scientific journals. *Scientometrics*, 35, 283–291.

Moed, H.F., and Van Leeuwen, T.N. (1996). Impact factors can mislead. *Nature*, 381, 186.

Moed, H.F., Van Leeuwen, Th. N., and Reedijk, J. (1998), A new classification system to describe the ageing of scientific journals and their impact factors. Journal of Documentation, 54, 387–419.

Moed, H.F. (2005). *Citation Analysis in Research Evaluation*. Dordrecht (Netherlands): Springer. ISBN 1-4020-3713-9, 346 pp.

Pinski, G., and Narin, F. (1976). Citation influence for journal aggregates of scientific publications: theory, with application to the literature of physics. *Information Processing and Management,* 12, 297–312.

Pudovkin, A.I., and Garfield, E. (2004). Rank-normalized impact factor. A way to compare journal performance across subject categories. Paper presented at the ASIST meeting, November 17, 2004. Available at: http://www.garfield.library.upenn.edu/papers/ asistranknormalization2004.pdf.

Reedijk, J., and Moed, H.F. (2008). Is the impact of journal impact factors decreasing? Journal of Documentation 64, 183-192.

Rousseau, R. (1988), Citation distribution of pure mathematics journals. In: L. Egghe, R. Rousseau (Eds), Informetrics 87/88, Elsevier Science Publishers B.V., pp. 249–260.

SCIMAGO. Description of Scimago Journal Rank Indicator. Retrieved 15 October 2009 from http://www.scimagojr.com/SCImagoJournalRank.pdf

Seglen, P.O. (1994). Causal relationship between article citedness and journal impact. *Journal of the American Society for Information Science*, 45, 1–11.

Seglen, P.O. (1997). Why the impact factor of journals should not be used for evaluating research. *British Medical Journal,* 314, 498–502.

Sen, B.K. (1992). Documentation Note: Normalised impact factor. *Journal of Documentation*, 48, 318–325.

Small, H., and Sweeney, E. (1985). Clustering the Science Citation Index using co-citations, I: A comparison of methods. *Scientometrics*, 7, 393–404.

Stringer, M.J., Sales-Pardo, M., and Amaral, L.A.N. (2008) Effectiveness of Journal Ranking Schemes as a Tool for Locating Information. PLoS ONE 3(2): e1683. doi:10.1371/journal.pone.0001683

Van Leeuwen, T.N., and Moed, H.F. (2002). Development and application of journal impact measures in the Dutch science system. Scientometrics , 53, 249-266.

West, J., Althouse, B., Rosvall, M., Bergstrom, T., and Bergstrom, C. (2008). Eigenfactor: Detailed methods. Retrieved April 24, 2008, from www.eigenfactor.org/ methods.pdf.

Zitt, M., and Small, H. (2008). Modifying the journal impact factor by fractional citation weighting: The audience *factor. Journal of The American Society for Information Science and Technology*, 59, 1856-1860.